\documentclass[prd,preprint,aps,dvips,showpacs]{revtex4}
\usepackage[english]{babel}
\usepackage{amscd}
\usepackage{epsfig}
\usepackage{tabularx}
\usepackage{graphicx}
\usepackage{latexsym}
\usepackage{amsmath}
\usepackage{amsfonts}
\usepackage{amssymb}
\usepackage[latin1]{inputenc}
\usepackage{times}
\usepackage[T1]{fontenc}
\usepackage{latexsym}
\usepackage{graphics}
\usepackage{verbatim}
\usepackage[absolute]{textpos}
\usepackage{wrapfig,times}
\usepackage{amsthm}
\usepackage{setspace}

\newcommand{\be}{\begin{equation}}
\newcommand{\ee}{ \end{equation}}
\newcommand{\ben}{\begin{eqnarray}}
\newcommand{\een}{\end{eqnarray}}

\begin{document}

\title{Fibonacci Oscillators in the Landau Diamagnetism problem}

\author{ André A. Marinho$^{a}$, Francisco A. Brito$^{b}$, Carlos Chesman$^{a}$}

\affiliation{$^{a}$ Departamento de Física Teórica e Experimental, Universidade Federal do Rio 
Grande do Norte, 59078-970 Natal, RN, Brazil. 
\\
$^{b}$ Departamento de Física, Universidade Federal de
Campina Grande, 58109-970 Campina Grande, Paraiba, Brazil
}
\date{\today}

\begin{abstract} 

We address the issue of the Landau diamagnetism problem via $q$-deformed algebra of Fibonacci oscillators through its generalized  sequence of two real and independent deformation parameters $q_1$ and $q_2$. 
We obtain $q$-deformed thermodynamic quantities such as internal energy, number of particles,  magnetization and magnetic susceptibility which recover their usual form in the degenerate limit $q_1^2 + q_2^2$=1.

\end{abstract}

\pacs{02.20-Uw, 05.30-d, 75.20-g}

\maketitle


\section{Introduction}

The Landau diamagnetism problem continues to play a role in several issues of many physical systems and has strong relevance today 
\cite{matt,kit,ste,van,lan}. The diamagnetism can be used as an illustrative 
phenomenon that plays essential role in quantum mechanics on the surface, the perimeter, and the dissipation of statistical 
mechanics of non-equilibrium. 

In this paper, we are interested in investigating this phenomenon in $q$-deformed algebra in order to understand impurities effects in, for
example, magnetization and susceptibility. The magnetic susceptibility is an intrinsic characteristic of a material and its identity is related to the atomic 
and molecular structure. In Ref.~\cite{omer} it was performed the calculation of susceptibility for electrons moving in a uniform 
external magnetic field, developing Landau diamagnetism, by applying the nonextensive Tsallis statistics \cite{tsa1,lav6,epb,lig}, 
which is a strong candidate for solving problems where the standard thermodynamics is not applicable --- see also Ref.~\cite{oze} for a
similar study using another method. Of course, other noncommutative deformations can be applied, for example $q$-deformation via Jackson derivative (JD) \cite{bri1}.

The study of quantum groups and quantum algebras has attracted great interest in recent years, stimulated intense research 
in various fields of physics \cite{bie,fuc}, taking into account a range of applications, covering cosmology and 
condensed matter, e.g. black  holes, fractional quantum Hall effect, high-temperature (high-T$_c$) superconductors \cite{wil}, 
rational field theories, noncommutative geometry, quantum theory of super-algebras and so on \cite{chai}. Furthermore, 
statistical and thermodynamic properties by studying  $q$-deformed physical  systems have been intensively investigated in the 
literature \cite{mac,flo,lav1,nar,crl1,mou,amg2,ams,zeng,bon2,vak,col,beh,su,liu}. 

Another important discussion is about the main reasons to consider {\it two} deformation parameters  in some different physical 
applications. Starting from the generalization of  the $q$-algebra \cite{jak},  in Ref.~ \cite{arik} it was 
generalized the Fibonacci sequence, which is a well-known linear combination where the third number is the sum of two predecessors 
and so on. Here, the numbers are in that sequence of generalized Fibonacci oscillators, where new parameters ($q_1,q_2$) are introduced
\cite{arik,amg1,aba1}.  They provide a unification of quantum oscillators 
with quantum groups, keeping the degeneration property of the spectrum invariant under the symmetries of the quantum group. 
The quantum algebra with two deformation parameters may have a greater flexibility when it comes to application in 
the concrete phenomenological physical models \cite{dao,gong}, and may increase interest in physical applications.

The paper is organized as follows. In Sec.~(\ref{alg}) we intrduce the $q$-deformed algebra. In Sec.~(\ref{fib}) we develop the 
($q_1, q_2$)-deformed Landau diamagnetism problem and in the Sec.~(\ref{con}) we make our final comments.

\section{Fibonacci oscillators algebra}
\label{alg}

We consider a system of generalized oscillators now entering two parameters in statistical distribution function, whose 
energy spectrum may be determined by Fibonacci's generalized sequence \cite{arik,amg1,aba1}. This will establish a statistical system 
depending on the deformation parameters ($q_1,q_2$), allowing us to calculate the thermodynamic quantities in the limit of high 
temperatures. 

The $q$-deformed quantum oscillator is now defined by the Heisenberg algebra in terms the annihilation and creation 
operators in $c$, $c^{\dagger}$, respectively, and the number operator $N$ \cite {lav1,aba1}, as follows
\be c_i c_{i}^\dagger - Kq_1^{2}c_{i}^\dagger c_i = q_2^{2n_i}\qquad\mbox{e}\qquad c_i c_{i}^\dagger - 
Kq_2^{2}c_{i}^\dagger c_i = q_1^{2n_i},\ee
\begin{equation}[N,c^{\dagger}] = c^{\dagger}, \qquad\qquad [N,c] = -c ,\end {equation}
where $K = \pm 1$, stands for bosons and fermions, respectively. In addition, the operators also obey the relations 
\begin{eqnarray}\label{e30.5}\;\;\qquad c^\dagger c=[N],\;\;\qquad cc^{\dagger} = [1+KN], \end {eqnarray}
\be [1+Kn_{i,q_1,q_2}] = Kq_1^{2}[n_i]+q_2^{2n_i},\;\quad\mbox{or}\quad\; [1+Kn_{i,q_1,q_2}] = Kq_2^{2}[n_i]+q_1^{2n_i}.\ee
The Fibonacci \textit{basic number} is defined as \cite{arik}
\be \label{e49}[n_{i,q_1,q_2}] = c_{i}^\dagger c_{i} = \frac{q_2^{2n_i}-q_1^{2n_i}}{q_2^2-q_1^{2}},\ee
The $q$-Fock space spanned by the orthornormalized eigenstates $|n \rangle$ is constructed according to 
\begin{eqnarray} {|n\rangle} =\frac{(c^{\dagger})^{n}} {\sqrt{[n]!}}{|0\rangle},\qquad\qquad c{|0\rangle}=0 ,\end {eqnarray}
The actions of $c$ e $c^{\dagger}$ and $N$ on the states $|n\rangle$ in the $q$-Fock space are known to be 
\begin{equation} c^{\dagger}{|n\rangle} = [n+1]^{1/2} {|n+1\rangle},\end{equation}
\begin{equation} c{|n\rangle} = [n]^{1/2} {|n-1\rangle},\end{equation}
\begin{equation} N{|n\rangle} = n{|n\rangle}.\end{equation}
To calculate the $q$-deformation statistical occupation number, we begin with the Hamiltonian of $q$-deformed noninteracting oscillators 
(bosons or fermions) \cite{chai},
\be {\cal H}_{q_1,q_2} = \sum_{i}{(\epsilon_i-\mu_{q_1,q_2})}{N_i},\ee
where $\mu_{q_1,q_2}$ is the {($q_1,q_2$)}-deformed chemical potential. It should be noted that this Hamiltonian is a two-parameter 
deformed Hamiltonian and depends implicitly on the deformation parameters $q_1$ and $q_2$, since the number operator is deformed 
via Eq.~(\ref{e49}).

The mean value of the ($q_1,q_2$)-deformed occupation number can be calculated by

\be [n_i]\label{e34}\equiv \langle[n_i]\rangle = \frac{tr(\exp(-\beta{\cal H}) 
c_{i}^\dagger c_{i})}{\Xi},\ee
\ben\label{e50}[n_{i,q_1,q_2}]=\frac{z'\left(\exp(\beta\epsilon_i)-z'\right)}
{\left(\exp(\beta\epsilon_i)-q_2^2z'\right)\left(\exp(\beta\epsilon_i)-q_1^2z'\right)},\een
where $z_{q_1,q_2}=\exp(\beta\mu_{q_1,q_2})$ is the fugacity of the system, and we shall use the notation $z_{q_1,q_2}=z'$. 
When {$q_1=q_2=1$}, we find the usual form
\ben \label{e50.0}n_i=\frac{1}{z^{-1}\exp{(\beta\epsilon_i)}-1}.\een
In the present application of the Fibonacci oscillators, we are interested in obtain new $(q_1,q_2)$-deformed thermodynamic quantities such as internal energy, magnetization, 
and magnetic susceptibility for the high-temperature case, i.e. the limit ($z\ll 1$).
\section{Fibonacci oscillators in the Landau diamagnetism}
\label{fib}

To explain the phenomenon of diamagnetism, we have to take into account the interaction between the external magnetic field 
and the orbital motion of electrons. Disregarding the spin, the Hamiltonian of a particle of mass \textit{m} and charge \textit{e} 
in the presence of a magnetic field \textbf{H} is given by the expression \cite{sal}
\begin{equation} \label{e52}{\cal H} = \frac {1}{2m} \left({p}-\frac{e}{c}{\bf A}\right) ^2, \end{equation}
where \textbf{A} is the vector potential associated with the magnetic field \textbf{H} and \textit{c} is the speed of light 
in $CGS$ units. Let us start to formalize the statistical mechanical problem by using the grand partition function with 
the parameters $q_1$ and $q_2$ inserted through Eq.~(\ref{e50}), in the form
\ben \label{e54} \ln\Xi &=& -K\frac{2eHL^2}{hc}\displaystyle\sum_{n}^{\infty}\frac{L}{2\pi}
\displaystyle\int_{-\infty}^{\infty}dk_{z}
\frac{1}{(q_1^2-q_2^2)}\Bigg\{\ln{\Big[1-Kz'q_1^2\exp(-\beta\epsilon)\Big]}(q_1^{-2}-1)+\nonumber\\
&+&\ln{\Big[1-Kz'q_2^2\exp(-\beta\epsilon)\Big](1-q_2^{-2})]}\Bigg\}, \een
where $k_{z} = -\infty,\cdots,\infty$, $\epsilon=\frac{\hbar k_{z}^2}{2m}+\hbar\omega\left(n+\frac{1}{2}\right)$, 
$\omega=\frac{eH}{mc}$. However, our study is focused on the analysis of diamagnetism 
in the limit of high temperatures $(z'\ll 1)$. Thus, performing the sum and integrals, we find the partition function is written as follows
\ben \label{e62}\ln\Xi=\frac{z'K^2HC_1}{\sinh(\gamma)}+\frac{z'^2K^3HC_2Q}{2\sinh(2\gamma)},\;\;\qquad \mbox{where}\qquad 
C_1=\frac{eL^3}{2\pi hc\lambda}, C_2=\frac{eL^3}{2\pi\sqrt{2}hc\lambda}\een
being  $\lambda=\frac{\hbar}{(2\pi m\kappa_B T)^{\frac{1}{2}}}$ the thermal wavelength, $\gamma=\beta\mu_{B}H$ and $Q=q_1^2+q_2^2-1$. 

We note that Eq.~(\ref{e62}) shows the ($q_1,q_2$)-deformation in the second term. In the first order does not appear $q$-deformation. It appears after considering at least the second order. Notice, however, the case 
($q_1=q_2=1$), { as expected, does recover the underformed thermodynamic quantities up to a second order correction which  are usually disregarded. On the other hand, taking computation up to second order corrections is necessary to get the effects of the $q$-deformation and as a consequence only in the unit circle on the ($q_1,q_2$)-space, i.e., 
\ben\label{q-circle}
q_1^2+q_2^2=1
\een
the deformation ceases}. The case in Eq.~(\ref{q-circle}) shows an interesting degeneration on the  ($q_1,q_2$)-space. Deformations show up as $q_1^2+q_2^2<1$ or $q_1^2+q_2^2>1$.  In former case appears the possibility of finding some unexpected negative thermodynamic  quantities such as negative specific heat. In the following we shall consider the latter case to calculate the $q$-deformed thermodynamic quantities of interest in the present study.

\subsection{($q_1,q_2$)-deformed thermodynamic quantities}
\label{nec}

We obtain the number of particles $N$ by setting,
\ben\label{e63}N=z'\frac{\partial}{\partial z'}\ln{\Xi}=\frac{z'K^2HC_1}{\sinh(\gamma)}+\frac{z'^2K^3HC_2Q}
{2\sinh(2\gamma)}.\een

We determine the internal energy, and we can write it in terms of $N$, in the form
\ben \label{e64} U=-\frac{\partial}{\partial\beta}\ln\Xi
=\frac{N\mu_B H\Big[C_1\coth(\gamma)\sinh(2\gamma)+z'KC_2\sinh(\gamma)\coth(2\gamma)Q\Big]}
{C_1\sinh(2\gamma)+z'KC_2\sinh(\gamma)Q}.\een

\begin{figure}[!htb]
\centerline{
\includegraphics[{angle=90,height=9.0cm,angle=270,width=9.0cm}]{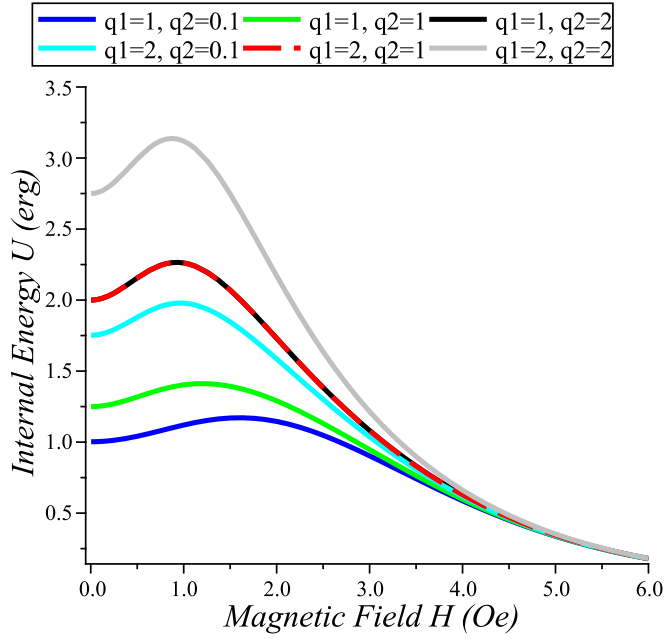}
}\caption{\small{($q_1,q_2$)-deformed internal energy as a function of magnetic field \textbf{H} for several choices of $q_1$ and $q_2$.}}\label{gráficos 1}
\end{figure}

In Fig.~\ref{gráficos 1} we have the behavior of internal energy $U$ as a function of the magnetic field \textbf{H} and for some values 
of $q_1$ and $q_2$ - see caption. We note that all the curves have different maximum peaks (depending on the values adopted for $q_1$ 
and $q_2$), for small magnetic  field $\textbf{H}$. The curves exhibit the same behavior asymptotically. We also have proven the symmetry 
between the oscillators, i.e. when $q_1=1$ and $q_2=2$ (black curve) and when $q_1=2$ and $q_2=1$ (red curve), they overlap. An expected 
effect due to the symmetry of $q_1$ and $q_2$ defined in $Q$.

The grand potential $\phi$ is determined as 
\be \label{e66.1}\phi=-\frac{1}{\beta}\ln\Xi=-\left(\frac{z'K^2HC_1}{\beta\sinh(\gamma)}+\frac{z'^2K^3HC_2Q}
{2\sinh(2\gamma)}\right).\ee

To determine the magnetization, we carried out the thermodynamic derivative by using Eq.~({\ref{e66.1}), that gives 
\ben \label{e67}M=-\frac{\partial\phi}{\partial H}=\frac{z'C_1K^2\left(1-\gamma\coth(\gamma)\right)}{\beta\sinh(\gamma)}-
\frac{z'C_2K^3Q\left(1-\gamma\coth(2\gamma)\right)}{2\beta\sinh(2\gamma)}.\een

We can also eliminate the chemical potential through the number of particles $N$ and insert the Langevin functions
\be \label{e67.2}{\cal L}(\gamma) = \coth(\gamma)-\frac{1}{\gamma},\qquad {\cal L}(2\gamma) = \coth(2\gamma)-\frac{1}{2\gamma},\ee
to rewrite the magnetization as 
\be \label{e67.5} M= -\frac{N\mu_B\Big[C_1\sinh(2\gamma){\cal L}(\gamma)+z'KC_2\gamma\sinh(\gamma){\cal L}(2\gamma)\Big]}
{C_{1}\sinh(2\gamma)+z'KC_{2}\sinh(\gamma)Q}.\ee

The results obtained for the deformed magnetization are very interesting, because we can compare it with experimental results 
obtained for superconducting materials (which are perfect diamagnetic materials) as a function of temperature variation \cite{gon}, in order to strength the
understanding of the $q$-deformation as a factor of impurity. In these references \cite{gon} one was found that the minimum of magnetization deepens as temperature or pressure decreases.

In Fig.~\ref{gráficos 2} we have the magnetization curves $(M)$ versus magnetic field (\textbf{H}) for some values of $q_1$ and $q_2$, 
and we note that some observations made for internal energy such as oscillators symmetry are also valid for the magnetization, as expected.  Notice that the minimum of magnetization deepens 
as $q$-deformation {\it increases}. This means that increasing temperature or pressure we may diminish the effects of disorders or impurities of the system. This explains why we should reduce the
deformation parameters until they assume the underformed degenerate case $q_1^2+q_2^2=1$.
\begin{figure}[htb]
\centerline{
\includegraphics[{angle=90,height=9.0cm,angle=270,width=9.0cm}]{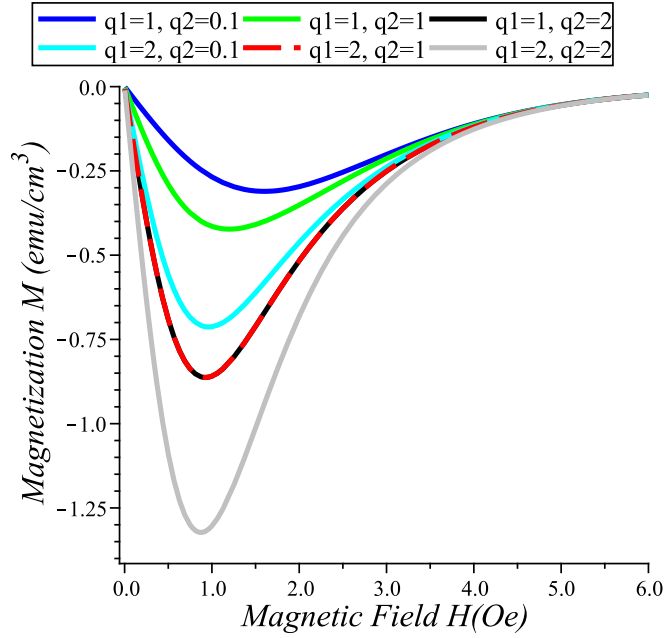}
}\caption{\small{($q_1,q_2$)-deformed magnetization as a function of magnetic field \textbf{H} for several choices of $q_1$ and $q_2$.}}\label{gráficos 2}
\end{figure}

Now, computing the susceptibility reads, 
\ben \label{e67.9}\chi&=&\frac{\partial M}{\partial H}
=\frac{N\beta\mu_B^2}{C_1\sinh(2\gamma)+z'KC_2\sinh(\gamma)Q}\Bigg[C_1\sinh(2\gamma)\Big(2\coth(\gamma)
{\cal L}(\gamma)-1\Big)+\nonumber\\&+& 2z'C_2\sinh(\gamma)Q\Big(2\coth(2\gamma){\cal L}(2\gamma)-1\Big)\Bigg].\een

In the limit of weak fields $\gamma\ll 1$ we have the leading term 
\be M=-\frac{2N\mu_B\sinh(\gamma)\cosh(\gamma)(C_1+z'C_2KQ)}{3(2C_1\cosh(\gamma)+z'KC_2Q)},\ee
and thus, we have the susceptibility in zero field 
\be \chi_0=-\frac{2\mu_B z'K^2(C_1+z'C_2KQ)}{3(2C_1+z'KC_2Q)}.\ee
\section{Conclusions}
\label{con}

As in our previous works \cite{bri1,bri2,bri3}, which we have shown that the $q$-parameter  is associated with
impurities in a sample, in particular diamagnetic materials, as in  the present study, we put forward new results to 
strength this interpretation of the $q$-deformation. 

In this work, we expand the application of $q$-calculation through two deformation parameters  ($q_1,q_2$), known as 
Fibonacci oscillators. We work in the limit of high temperatures (`dilute gas' $z\ll 1$), and a  
($q_1,q_2$)-deformed partition function.  In first order the results reported in the literature \cite{sal,patt}, are recovered. However, the $q$-deformation takes place  at  second order
for non-degenerate case $q_1^2+q_2^2>1$.

We note that the in the obtained results  were found several interesting behaviors by just varying the values of $q_1$ and $q_2$.
Of course, we performed a theoretical application, and it allows various assumptions. By comparing these results 
with similar experimental curves, one could understand how impurities could be entering into a material that affects, e.g., superconductivity such its 
critical temperature increases, which would be of  great interest to whom that works with high T$_c$ superconductors --- see \cite{Brito:2012gp} for a recent alternative  
theoretical investigation on these type of superconductors whose structure can be extended via $q$-deformation in order to introduce impurities.
\acknowledgments

We would like to thank CNPq, CAPES, and PROCAD-CAPES, for partial financial support.

\end{document}